\documentclass[letterpaper]{article} 
\usepackage[draft]{aaai25}
\usepackage{times}  
\usepackage{helvet}  
\usepackage{courier}  
\usepackage[hyphens]{url}  
\usepackage{graphicx} 
\usepackage{natbib}  
\usepackage{caption} 
\usepackage{algorithm}
\usepackage{algorithmic}

\usepackage{xcolor} 
\newcommand{\KL}[1]{\textcolor{black}{#1}}
\newcommand{\UB}[1]{\textcolor{black}{#1}}

\usepackage{newfloat}
\usepackage{listings}
\DeclareCaptionStyle{ruled}{labelfont=normalfont,labelsep=colon,strut=off} 
\floatstyle{ruled}
\newfloat{listing}{tb}{lst}{}
\floatname{listing}{Listing}
\usepackage{booktabs} 
\usepackage{amsmath}

\title{Disaggregated Health Data in LLMs: Evaluating Data Equity \\in the Context of Asian American Representation}
\author {
    Uvini Balasuriya Mudiyanselage,
    Bharat Jayprakash,
    Kookjin Lee,
    K. Hazel Kwon
}
\affiliations {
    Arizona State University, Tempe, AZ, USA\\
    ubalasur@asu.edu, bjayprak@asu.edu, kookjin.lee@asu.edu, khkwon@asu.edu
}

\usepackage{bibentry}

\begin{document}

\maketitle

\begin{abstract}
  Large language models (LLMs), such as ChatGPT and Claude, have emerged as essential tools for information retrieval, often serving as alternatives to traditional search engines. However, ensuring that these models provide accurate and equitable information tailored to diverse demographic groups remains an important challenge. This study investigates the capability of LLMs to retrieve disaggregated health-related information for sub-ethnic groups within the Asian American population, such as Korean and Chinese communities. Data disaggregation has been a critical practice in health research to address inequities, making it an ideal domain for evaluating representation equity in LLM outputs. We apply a suite of statistical and machine learning tools to assess whether LLMs deliver appropriately disaggregated and equitable information. By focusing on Asian American sub-ethnic groups—a highly diverse population often aggregated in traditional analyses—we highlight how LLMs handle complex disparities in health data. Our findings contribute to ongoing discussions about responsible AI, particularly in ensuring data equity in the outputs of LLM-based systems.
\end{abstract}

%

\section{Introduction}

This study examines the ability of large language models (LLMs) to retrieve disaggregated health information. LLMs, such as GPT-3~\cite{brown2020language}, Claude~\cite{anthropic2023claude}, Perplexity~\cite{perplexity2023}, and many others, have transformed the way people seek information. These generative AI services have become widespread, serving as significant supplements—or even alternatives—to traditional web search-based information retrieval systems. Their content generation capabilities, combined with human-machine conversational interactivity, assist the broader public in accessing scientific information and knowledge with ease.

However, the proliferation of LLM-based systems has raised concerns about the risks associated with their use. As a result, the concept of ‘responsible AI’ has recently gained prominence in the development and deployment of AI services, emphasizing the need to incorporate ethical considerations into practice. Among the various rights-based principles underpinning responsible AI practices, this study focuses on ‘data equity.’ Specifically, it examines the extent of representation equity in the disaggregated health information retrieved by LLMs. Data disaggregation and data equity are inseparable. Without disaggregated data collection and analysis, data equity cannot be effectively integrated into policymaking. Conversely, without a commitment to data equity, the disaggregated assessment of underserved communities becomes impossible. By systematically reviewing ways in which LLMs generate disaggregated information, this paper evaluates the state of representation equity in publicly available health-related data concerning ethnic groups.

\section{Data Disaggregation and Data Equity in LLMs}

Policymakers and experts have recently reached a consensus that data equity is a necessary condition for addressing various contemporary equity issues in a multiethnic society like the U.S.  Notably, the White House recently launched the Equitable Data Working Group to emphasize the importance of data equity in policymaking: “Equitable data…allow for rigorous assessment of the extent to which government programs and policies yield consistently fair, just, and impartial treatment of all individuals. Equitable data illuminate opportunities for targeted actions that will result in demonstrably improved outcomes for underserved communities” \cite{WhiteHouse2022}. 

In consultation with 50 civil society and research organizations, the Working Group recommended \textit{data disaggregation} as one of its top priorities. Data disaggregation involves collecting, analyzing, and reporting data by purposefully accounting for underserved populations, ensuring the most marginalized groups are represented in the data used for policy decision-making. For instance, in the U.S., ``people of Middle Eastern and North African heritage and subgroups of Asian American, Native Hawaiian, and Pacific Islanders'' are poorly represented within the current demographic categories. This exclusion leaves them unseen in government statistics and masks critical inequities \cite{WhiteHouse2022}.

Disaggregated data practices can reveal disparities in various aspects of well-being. For example, among Asian American populations, Indian Americans had a median household income of \$123,700 in 2019, whereas Burmese Americans had a median income of \$44,400 ~\cite{Pew2021}. Approximately 25\% of Burmese Americans and 19\% of Hmong Americans live below the poverty line, compared to the national average of 10.5\% ~\cite{Pew2021}. Regarding educational attainment, while many Asian American subgroups have high levels of achievement, disparities persist. For example, 74\% of Indian Americans aged 25 and older hold at least a bachelor's degree, compared to only 15\% of Cambodian Americans and 14\% of Laotian Americans~\cite{Pew2019}. Additionally, only about 73\% of Pacific Islander students graduated from high school on time, compared to the national average of 86\% ~\cite{NCES2021}.

While these examples underscore the importance of data disaggregation in identifying and addressing inequities among underrepresented minority groups, its role in generating informational outputs for LLMs remains largely unexplored. In the context of LLMs, data equity issues are even more nuanced. In traditional population research, where survey data is created and used, data equity discussions typically focus on problems associated with data collection, such as sampling errors (due to difficulty accessing participants from underrepresented ethnic communities), non-response errors (stemming from a community’s distrust of institutional data collection, rooted in historical and cultural memories), and coverage errors (caused by structural factors like biased census districting) \cite{shimkhada2021capturing,sasa2022just}.

In contrast, the LLM-based decision system introduces additional layers of complexity. LLMs likely train itself by using \textit{not} raw data (e.g., survey responses) but instead publicly available secondary informational content (e.g., policy reports, news articles, research papers) derived from raw data. Consequently, biases inherent in the raw data are compounded by biases introduced during the creation of secondary sources. Furthermore, certain ``interesting'' findings may appear across multiple content forms and be repeatedly fed into the system (e.g., Pew survey results~\cite{Pew2019,Pew2021} cited in news and scholarly articles). Moreover, how algorithms prioritize and process these sources remains unclear. Finally, the generated outputs of LLMs can become looped back into their training phases, amplifying existing biases. As a result, data equity in LLMs permeates the entire data lifecycle—from input to algorithmic processing to output.

Given this complexity, this study aims to contribute to discussions on data equity in LLMs. Focusing on ``representation equity,'' \cite{stonier2023data}, we analyze the quality of disaggregated information retrieved by LLMs. This study centers on Asian Americans—a diverse racial population comprising numerous sub-ethnic groups—who stand to benefit significantly from improved data disaggregation practices. In terms of issue domain, we focus on health. Health policy is one of the key sectors that have shown the advancement in practicing data disaggregation. For example, when cancer rates are reported in an aggregated manner, the average rate for Asian women appears lower than that for non-Hispanic White women. However, disaggregated records reveal that Laotian women have cancer rates more than nine times higher than those of White women (WHIAANHPI, \cite{whitehouse2023National}). Similarly, disaggregated data show that Vietnamese American women have the highest rates of cervical cancer among all racial/ethnic groups~\cite{NCI2020}. Native Hawaiian and Pacific Islanders (NHPI), along with Indian and Pakistani Americans, have a higher prevalence of Type 2 diabetes (15\% and 12.6\%, respectively) compared to non-Hispanic Whites (7.6\%), while Chinese Americans exhibit a significantly lower rate (4.3\%) \cite{CDC2020}. \KL{A recent study \cite{movva2023coarse} also shows that using coarse race groups in U.S. healthcare data can mask important disparities among more specific subpopulations.}
By examining the quality of information retrieval in the relatively well-disaggregated domain, we identify the current status of representation equity of Asian American ethnic groups in LLM outputs. Based on the discussions, we proposes three research questions (RQs): 
\begin{itemize}
    \item RQ1: How is LLM's health information retrieval in the context of ethnic minority groups (dis)similar to that in the context of broader population? 
    \item RQ2: How diverse is health information represented in LLM outputs for various ethnic minority groups? 
    \item RQ3: How accurate is health information represented in LLM  outputs for various ethnic minority groups?  
\end{itemize}

\section{Data Collection}
In this study, we choose ChatGPT~\cite{brown2020language} by OpenAI~\cite{openai_api} as the platform for data collection for several key reasons. The primary reason is that ChatGPT has emerged as one of the most widely used LLMs globally, boasting a diverse and extensive user base. This widespread adoption ensures that its outputs reflect a broad range of real-world interactions, making it an ideal candidate for examining representation equity across diverse demographic groups. 
Also, OpenAI's API grants access to relatively recent models, including GPT-3.5, which exhibit advanced reasoning capabilities and a vast knowledge base. These features make the platform particularly well-suited for studying nuanced patterns in disaggregated data and addressing the complexities of representation equity. \KL{We focus on GPT-3.5 in this study because its strong reasoning abilities and broad adoption align with our research goals; we do not examine cross-model or cross-platform output consistency.}

The data collection process is as follows. 
We collect data using the OpenAI's API \cite{openai_api}, focusing on 21 sub-ethnic categories in Asian American population based on the U.S. Census Bureau breakdown \cite{census2022aanhpi} (See Table~\ref{tab:keywords}). 
\begin{table}[!b]
  \caption{Ethnicity listed in Census (alphabetically-ordered)}
  \label{tab:keywords}
  \begin{tabular}{p{0.95\columnwidth}}
    \toprule
    \multicolumn{1}{c}{\textbf{Ethnic categories}}  \\
    \hline
     Bangladeshi, Bhutanese, Burmese, Cambodian, Chinese, Filipino, Hmong, Indian, Indonesian, Japanese, Korean, Laotian, Malaysian, Mongolian, Nepalese, Okinawan, Pakistani, Sri Lankan, Taiwanese, Thai, Vietnamese\\
    \bottomrule
\end{tabular}
\end{table}
\KL{We use this Census-based categorization because it aligns with how ethnic subgroups are typically represented in the medical and public health domains. While we acknowledge that further subcategorization may be valuable—for example, distinguishing among regional or linguistic subgroups—such granularity is beyond the scope of this study due to the lack of standardized guidelines.}

The prompt used is: 
\begin{quote}
\textit{``Generate important health issues that I should be concerned about as a [demographic group]. Be faithful to all the provided demographic features I did not specify.''}
\end{quote}
To ensure a diverse and balanced dataset, we set the number of generations hyperparameter to 10 and loop it 100 times to gather 1,000 generations for each demographic group. Generating all 1,000 samples in one step risks producing repetitive, biased, or lower-quality responses due to the model's contextual and computational constraints.

Data is collected using the \textsc{GPT}-3.5 model (“gpt-3.5-turbo”) with default hyperparameters, except for setting the temperature to 1 to encourage a wider variety of responses. All data collection occurs between August and October 2024.

\section{Analysis}
In this section, we outline the methods employed to address the research questions and present the findings derived from our analysis. We conduct a comprehensive set of investigations across various levels of information granularity (word-level, topic-level, and health-condition-level), response granularity (aggregated samples and per-sample analysis), and both quantitative and qualitative approaches. By thoroughly examining these dimensions, we aim to assess patterns of representation equity, diversity, and accuracy in LLM outputs. The subsequent subsections provide detailed descriptions of the methods and findings for each level of analysis, aligned with our three research questions.

\subsection{RQ1: How is LLM’s health information retrieval in the context of ethnic minority groups
(dis)similar to that in the context of broader population}

To answer RQ1, we employ multiple statistical and ML methods with various levels of granularity of the data, namely, health-condition-level, word-level and topic-level analyses. In all considered methods, we compute the distance from the reference group [Asian American] to each specific ethnicity. For health-condition-based analysis, we created a list of health conditions for each group, 
sort the list in descending order based on the frequencies and utilize two classical statistical metrics to measure the distance from the reference group by comparing the sorted list of a group to that of the reference group. 
We detail the extraction process in the methodology described under RQ3, where a more relevant analysis is performed. 
For the topic-level analysis, we employ the \textsc{BERTopic} API \cite{grootendorst2022bertopic} to measure the distance in an embedding space constructed through a topic modeling. After creating topic clusters, we perform correspondent analysis (CA) to compute an embedding space and measure a distance in that embedding space. 

\subsubsection{Methods}
We first introduce two statistical methods to compute distances between two ranked lists and then explain the details about the topic modeling process, followed by CA to create an embedding space for categorical variables (i.e., ethnic groups) in terms of topical similarities. 
\paragraph{\textbf{Health-condition-level analysis}}
We perform the Kendall-Tau Rank Correlation Coefficient and the Kendall-Tau Rank Distance. The Kendall--Tau rank correlation coefficient (\(\tau\)) measures the strength and direction of association between the ranked conditions in each ethnicity and the broadly defined `Asian American' group. It is defined as:
\begin{equation}
    \tau = \frac{n_c - n_d}{\binom{n}{2}}    
\end{equation}
where \(n_c\) denotes the number of concordant pairs (pairs ranked in the same order in both groups),  \(n_d\) denotes the number of discordant pairs (pairs ranked in different orders in the two groups), and 
\(\binom{n}{2}\) denotes Total number of possible pairs for \(n\) ranked items.  A \(\tau\) value of \(+1\) indicates perfect agreement, \(0\) indicates no correlation, and \(-1\) indicates perfect disagreement.

The Kendall-Tau rank distance quantifies the number of pairwise disagreements between the rankings of the two groups, which is defined as:
\begin{equation}
    d_\tau = n_d.  
\end{equation}
This distance focuses exclusively on the discordant pairs \(n_d\), with higher values indicating greater divergence between the rankings of the two groups. By examining both rank correlation and distance, we capture not only the similarity in the ordering of conditions between each ethnic group and the American group but also the magnitude of the differences.

\paragraph{\textbf{Topic-level analysis: BERTopic and Correspondence Analysis (CA)}}
We perform topic modeling using the \textsc{BERTopic} API \cite{grootendorst2022bertopic}, a Transformer-based technique that generates human-interpretable results by embedding documents into vector representations, reducing their dimensions, clustering them, and identifying latent topics based on the most representative words in each cluster. For this process, we utilize Sentence-Transformer \cite{Reimers2019SentenceBERT} for document embedding, UMAP \cite{McInnes2018UMAP} for dimensionality reduction, and HDBSCAN \cite{McInnes2017HDBSCAN} for clustering. For details, we refer readers to Appendix~\ref{app:bertopic}. 

To complement the topic modeling, we take a further step by performing  Correspondence Analysis (CA), a multivariate statistical technique used to analyze the relationships between two categorical variables in a contingency table by transforming rows and columns into a lower-dimensional space. CA is conducted for all 21 sub-ethnic Asian American categories, where ``Asian American'' is included as a reference group. In the analysis, the cluster composition of Asian Americans is compared relative to the 21 sub-ethnic Asian American categories. This approach enables us to explore deviations within subcategories more comprehensively, providing deeper insights into variations compared to the reference Asian American group. 

Before performing CA, we derive multiple topics for each demographic group from the previous topic modeling results, which makes interpreting all the topics for over 20 demographics challenging. To further assess the distances to reference groups and identify topics most closely associated with specific demographics, we proceed as follows. First, we combine all topic embeddings across demographics and perform clustering by reducing their dimensionality using UMAP. We then apply two clustering algorithms---HDBSCAN and DBSCAN---across various dimensions (ranging from 2D to 8D). To evaluate the effectiveness of these clustering methods, we use metrics such as the Silhouette Score and the Davies--Bouldin Index. We then create a single vector for each demographic by calculating the proportion of topics within that demographic that belong to each cluster such that $[\rho_1^{g}, \rho_2^{g}, \ldots]$, where $\rho_i^{g}$ denotes the proportion of topics in cluster $i$ for the demographic group~$g$. 

We then perform CA by first constructing a contingency table, where the rows represent clusters, and the columns represent demographic groups and by applying singular value decomposition to the table matrix to extract orthogonal components for each item in the table.

After performing topic modeling and CA, we obtain two embedding spaces: one generated from the topic model and another from the CA results, where in both spaces ethnic subgroups and topics are represented as numerical vectors, To measure the distance between each ethnicity and the reference group, we use two metrics for the topic model embeddings: Jensen--Shannon divergence (JSD) and cosine distance. For the CA embeddings, we use Euclidean distance. JSD quantifies the difference between two probability distributions, cosine distance measures the angular difference between two vectors, and Euclidean distance captures the straight-line distance between two points in the vector space (see Appendix~\ref{app:CA} for details).

\subsubsection{Analysis}
Table~\ref{tab:kendall_tau} shows the results of applying the various distance measures described above.  Table~\ref{tab:kendall_tau} highlights significant variations across sub-ethnic Asian American groups. On the one hand,  Chinese, Taiwanese, and Korean closely aligned with the reference group (Asian American). In contrast, other ethnicities like Filipino, Burmese, and Sri Lankan show substantial divergence. Geographic and cultural proximity might influence these patterns, with East Asians generally clustering closer to the reference group and Southeast Asians and South Asians showing more variability. In the following we provide more detailed analysis: 

\begin{table}[htbp]
\centering
\footnotesize 
\begin{tabular}{l|c|c|c|c|c}
\toprule
\textbf{Demography} & \textbf{$\tau$ ($p$-value)} & \textbf{$d_\tau$} & CD & JSD & ED \\ 
\midrule
Bangladeshi  & 0.51 (0.0015) & 0.24 & 0.24 & 0.09 & 0.26 \\
Bhutanese    & 0.32 (0.0587) & 0.34 & 0.24 & 0.27 & 1.26 \\
Burmese      & 0.37 (0.0249) & 0.31 & 0.25 & 0.33 & 1.30 \\
Cambodian    & 0.53 (0.0033) & 0.23 & 0.19 & 0.31 & 1.29 \\
Chinese      & 0.77 (0.0000) & 0.11 & 0.10 & 0.07 & 0.33 \\
Filipino     & 0.44 (0.0071) & 0.28 & 0.17 & 0.30 & 2.46 \\
Hmong        & 0.38 (0.0413) & 0.31 & 0.24 & 0.32 & 1.29 \\
Indian       & 0.22 (0.1950) & 0.39 & 0.12 & 0.11 & 1.31 \\
Indonesian   & 0.44 (0.0054) & 0.28 & 0.20 & 0.21 & 0.70 \\
Japanese     & 0.74 (0.0000) & 0.13 & 0.13 & 0.11 & 1.31 \\
Korean       & 0.64 (0.0001) & 0.18 & 0.13 & 0.06 & 0.24 \\
Laotian      & 0.42 (0.0103) & 0.29 & 0.18 & 0.26 & 1.07 \\
Malaysian    & 0.44 (0.0110) & 0.27 & 0.20 & 0.13 & 0.41 \\
Mongolian    & 0.41 (0.0143) & 0.29 & 0.19 & 0.22 & 0.79 \\
Nepalese     & 0.37 (0.0200) & 0.31 & 0.22 & 0.18 & 0.91 \\
Okinawan     & 0.49 (0.0036) & 0.25 & 0.24 & 0.28 & 2.30 \\
Pakistani    & 0.49 (0.0028) & 0.25 & 0.21 & 0.15 & 1.60 \\
Sri Lankan   & 0.30 (0.0640) & 0.34 & 0.24 & 0.21 & 2.04 \\
Taiwanese    & 0.77 (0.0000) & 0.11 & 0.11 & 0.09 & 0.25 \\
Thai         & 0.59 (0.0002) & 0.20 & 0.16 & 0.23 & 1.10 \\
Vietnamese   & 0.48 (0.0031) & 0.26 & 0.20 & 0.21 & 1.01 \\
\bottomrule
\end{tabular}
\vspace{1mm}
\caption{Kendall-Tau correlation coefficient ($\tau$), Kendall-Tau rank distance ($d_\tau$), cosine distance (CD) and Jensen-Shannon divergence (JSD) in the topic embedding space, and Euclidean distance in the CA embedding space (ED). All metrics are measured against the reference group, Asian American.}
\label{tab:kendall_tau}
\end{table}

\begin{figure}[tbp]
    \centering
    \includegraphics[width=0.8\columnwidth]{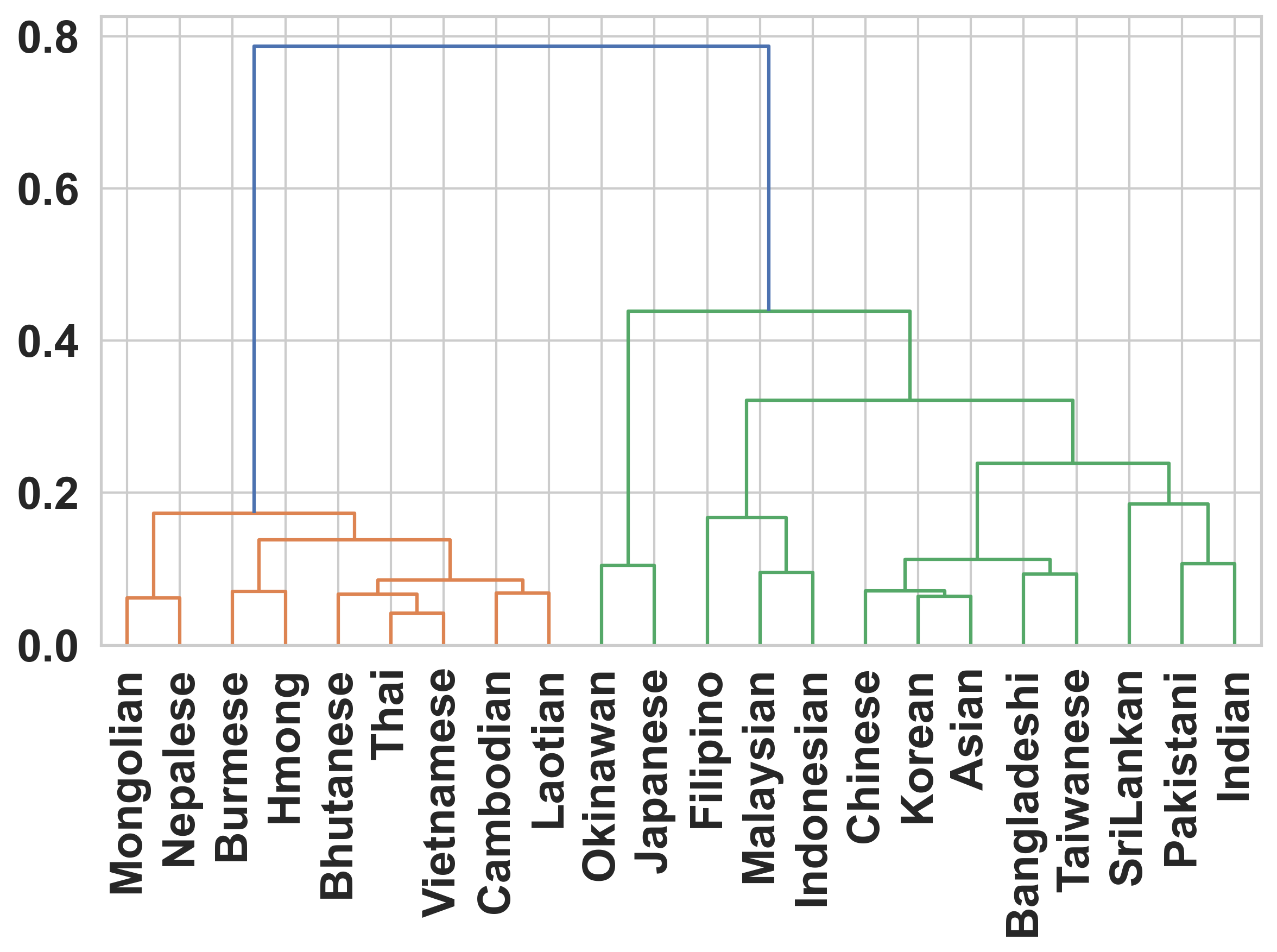}
    \caption{[Topic modeling] Hierarchical clustering of subethnic Asian American demographic groups based on the Jensen--Shannon divergence of topic cluster distributions.}
    \label{fig:BERTopic_hierachical_clustering}
\end{figure}

\begin{figure}[tbp]
    \centering
    \includegraphics[width=0.8\columnwidth]{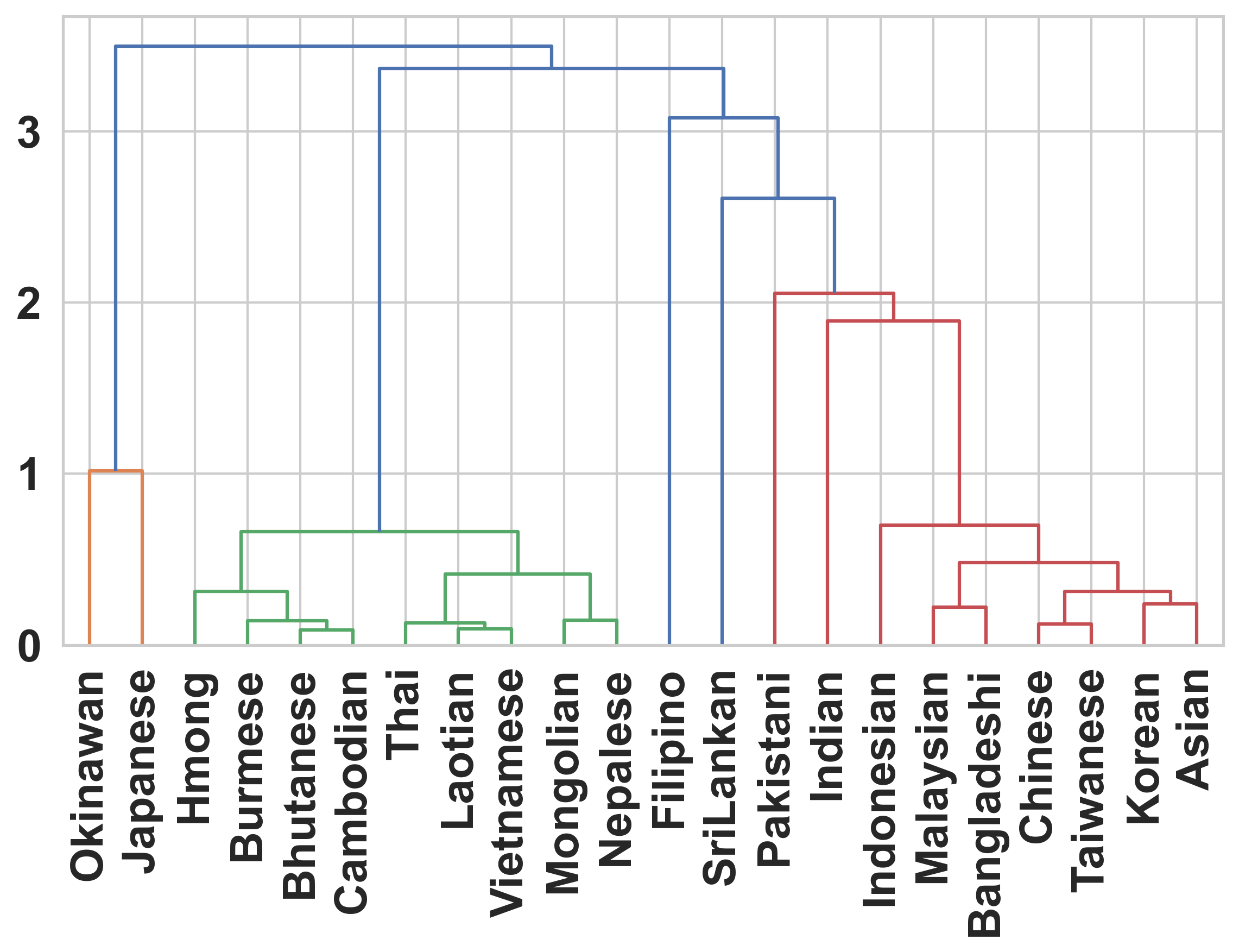}
    \caption{[CA] Hierarchical clustering of subethnic Asian American demographic groups based on their similarity in the embedding space, measured in the Euclidean distance.}
    \label{fig:MCA_hierachical_clustering}
\end{figure}

\begin{itemize}
    \item Low distance groups: Chinese, Taiwanese, and Korean show the smallest distances in all measures, indicating their close alignment with the reference group, `Asian American'. Chinese has the lowest overall distance suggesting that it is the most representative; Taiwanese and Korean also exhibit minimal distance across all metrics, highlighting their proximity in characteristics to the reference group. These group exhibit higher Kendall-Tau rank correlation $\tau$ and lower Kendall-Tau rank distance $d_\tau$ because they may have similar cultural, social, or historical factors that align their rankings more closely with the broader ``Asian American'' category.
    \item High distance groups: Filipino, Burmese, and Sri Lankan have some of the largest distances across all metrics, reflecting significant divergence from the reference group. Filipino stands out with the highest ED (2.4595) and a high JSD (0.3049), indicating distinct differences in both CA and topic embedding spaces. Burmese and Sri Lankan also reflect notable divergence in all metrics. Also, the groups of a higher degree of divergence in cultural or social factors, like Bangladeshi, Pakistani, or Sri Lankan, might show lower $\tau$ and higher $d_\tau$, suggesting that their rankings differ more significantly from the ``Asian American'' group.
    \item Regional clustering: East Asian groups like Chinese, Taiwanese, Japanese, and Korean tend to cluster near the reference group, as reflected by their lower values across all metrics. Southeast Asians  groups such as Burmese, Cambodian, Laotian, and Vietnamese generally show higher distances (e.g., ED around 1.0–1.3) and moderate-to-high JSD values ($>0.2$), indicating regional differentiation. South Asians  groups like Indian, Bangladeshi, Pakistani, and Sri Lankan exhibit moderate to high divergence, particularly in $d_\tau$ and ED, reflecting distinct cultural and demographic characteristics. Figures~\ref{fig:BERTopic_hierachical_clustering} and ~\ref{fig:MCA_hierachical_clustering} depict the hierarchical clustering of subethinic groups based on the distance measured in JSD (in the topic cluster distribution generated by BERTopic) and Euclidean distance (in the embedding space generated by CA). Figures again confirm that the hierarchical clusters show patterns that are aligned with geographical distributions of each subethnicities. 
    \item Some inconsistent entries: Bhutanese, Filipino, and Okinawan show some inconsistency between different measures. Bhutanese and Okinawan populations are relatively smaller compared to larger groups like Chinese or Indian Americans, which might make their rankings more susceptible to biases or inconsistencies in different measures. Filipinos are a large and diverse group, but their representation might vary depending on region, socioeconomic status, or generational factors, leading to inconsistent rankings. 
\end{itemize}

BERTopic modeling combined with CA further provides some interpretiblity on which topic clusters attribute to the similarity of the subgroup to the reference group. To identify the clusters most strongly associated with each demographic group, we use a modified metric of ``relationship strength'' based on the CA results. After transforming the demographic groups and clusters into a lower-dimensional space, we determine the optimal  number of dimensions that explain 80\% of the cumulative variance, based on the eigenvalues of the CA decomposition. We refer readers to Appendix~\ref{app:CA} for more detailed description of the method and the full results tables. Each topic is given an arbitrary number only for indexing purposes. 

The results of topic-level analysis align with the health-condition -level analysis. East Asian populations (Korean, Taiwanese, and Chinese) exhibit the closest distance to the reference group, where the most prevalent topics correspond to concerns on vision and age-related conditions (i.e., topic \#4: \{exams, eye, vision, glaucoma, cataracts, ...\}) and concerns on cardiovascular and diabetes (i.e., topic \#3: \{kidney, obesity, hypertension, type, diabetes \}. The next closest group consists of South Asians (Bangladeshi, Malaysian, Indonesian, Indian, Pakistani, and Sri Lankan), exhibiting a shared concern on mental health and acculturation (i.e., topic \#27: underreporting, anxiety, depression, mental, acculturation). The populations in the second group (Nepalese, Mongolian, ..., Hmong, marked in green color in Figure~\ref{fig:MCA_hierachical_clustering}) and the third group (Japanese and Okinawan, marked in orange color in Figure~\ref{fig:MCA_hierachical_clustering}) exhibit topics that are very distinctive from the closest groups. The second group's topics include ``oral hygiene issues'' (topic \#18) and ``access to healthcare'' (topic \#19)  and the third group's topics include ``obesity'' (topic \#30) and ``bone-related issue'' (topic \#11).

\subsection{RQ2: How diverse is health information represented in LLM outputs for various ethnic minority groups}

In this section, we explore the variability of responses within each subgroup to assess differences in how information is represented. Specifically, we analyze three metrics: the number of words (nWords), the number of distinct conditions (nConds), and a variability measure (Simpson's diversity) for responses within each subgroup. By comparing these metrics across subgroups and against the reference group, we aim to identify whether certain subgroups exhibit greater or lesser variability in their responses.

\begin{table*}[t]
\centering
\begin{tabular}{l|l|c|c|c|c|c|c}
\hline
& & \multicolumn{2}{c|}{\textbf{nWords}} & \multicolumn{2}{c|}{\textbf{nConds}} & \multicolumn{2}{c}{\textbf{Simpson's Diversity}} \\
\hline
\textbf{Group 1} & \textbf{Group 2} & \textbf{Mean Diff} & \textbf{P-Adj} & \textbf{Mean Diff} & \textbf{P-Adj} & \textbf{Mean Diff} & \textbf{P-Adj} \\
\hline
Asian & Bangladeshi & 14.4051 & 0.0000 & 0.3420 & 0.0001 & -0.0011 & 0.0022 \\
Asian & Bhutanese & -0.5748 & 1.0000 & -0.1955 & 0.5700 & 0.0004 & 0.9981 \\
Asian & Burmese & -4.5000 & 0.9031 & -0.8211 & 0.0000 & -0.0023 & 0.0000 \\
Asian & Cambodian & -6.0154 & 0.4380 & -0.4820 & 0.0000 & -0.0010 & 0.0599 \\
Asian & Chinese & -3.1825 & 0.9976 & 0.0819 & 1.0000 & 0.0007 & 0.5746 \\
Asian & Filipino & 17.1536 & 0.0000 & 0.2642 & 0.0276 & -0.0015 & 0.0000 \\
Asian & Hmong & -7.5244 & 0.0538 & -0.5863 & 0.0000 & -0.0013 & 0.0003 \\
Asian & Indian & 35.1425 & 0.0000 & 1.1801 & 0.0000 & -0.0002 & 1.0000 \\
Asian & Indonesian & -2.6033 & 0.9997 & -0.0461 & 1.0000 & -0.0007 & 0.4964 \\
Asian & Japanese & -4.4391 & 0.8661 & -0.3478 & 0.0002 & -0.0009 & 0.1091 \\
Asian & Korean & -5.7470 & 0.4685 & -0.4772 & 0.0000 & -0.0011 & 0.0031 \\
Asian & Laotian & -22.3485 & 0.0000 & -0.6604 & 0.0000 & -0.0006 & 0.7824 \\
Asian & Malaysian & 12.1419 & 0.0000 & -0.2386 & 0.0816 & -0.0033 & 0.0000 \\
Asian & Mongolian & 1.0826 & 1.0000 & -0.3365 & 0.0013 & -0.0008 & 0.2336 \\
Asian & Nepalese & 17.0852 & 0.0000 & -0.1794 & 0.6518 & -0.0022 & 0.0000 \\
Asian & Okinawan & 1.8414 & 1.0000 & -0.2602 & 0.0458 & -0.0007 & 0.3446 \\
Asian & Pakistani & 27.6519 & 0.0000 & 0.6483 & 0.0000 & -0.0004 & 0.9947 \\
Asian & SriLankan & 18.8457 & 0.0000 & 0.5054 & 0.0000 & -0.0002 & 1.0000 \\
Asian & Taiwanese & 5.2620 & 0.7304 & 0.0068 & 1.0000 & -0.0005 & 0.9910 \\
Asian & Thai & -7.5033 & 0.0936 & -0.3590 & 0.0005 & 0.0001 & 1.0000 \\
Asian & Vietnamese & -18.9926 & 0.0000 & -0.6267 & 0.0000 & -0.0012 & 0.0009 \\
\hline
\end{tabular}
\caption{Posthoc Tukey HSD test results for the number of words, the number of health conditions, and Simpson's Diversity.}
\label{tab:anova_combined}
\end{table*}

\subsubsection{Methods}
To examine the informational diversity in the output generated by LLM, we treat each response text as a unit of analysis. We clean the text corpus by removing formatting artifacts such as newlines, tabs, bullet points, and unnecessary punctuation. Using a regular expression pattern, we extract health-related conditions, which we also used in earlier analyses addressing RQ1. \KL{We refer readers to Appendix for the details of the regular expression we use.} Within each text, we calculate text-level metrics including 
number of words, number of unique health conditions, and Simpson's diversity score. We then compare these metrics across demographic groups. By comparing each metric, assess the overall information volume, variety of health-specific information,  and details in describing each identified health condition  (Simpson's diversity). To compute Simpson's diversity, we first consider health conditions as categories and the total length of sentences that contain the health condition word as counts. With this approach, the measurement is expected to indicate the variability in terms of how different health conditions are articulated (e.g., cancer could be described with 100 words while flu would be described with 20 words).

We use one-way analysis of variance (ANOVA)  to evaluate whether the mean of those metrics are significantly different across demographic groups. This approach partitions the total variance into between-group and within-group components to compute an F-statistic. The null hypothesis tested in ANOVA is that the mean values of the response-level variables are equal across all demographic groups, while the alternative hypothesis is that at least one group's mean differs from the others.

Following a significant ANOVA result, we perform post-hoc Tukey HSD (honestly significant difference) tests \cite{tukey_hsd} to identify specific group pairs with significant mean differences. Tukey HSD accounts for multiple comparisons, providing adjusted p-values and confidence intervals for pairwise mean differences. The null hypothesis for each pairwise comparison in Tukey HSD is that the mean values of the variable for the two groups being compared are equal, while the alternative hypothesis posits a significant difference between the two means.

\subsubsection{Analysis}
Table~\ref{tab:anova_combined} reports the results of post-hoc Tukey HSD tests on 
the number of words, the number of health conditions, and Simpson's Diversity. While some groups showed greater overall informational volume than Asian American reference (nWords), the results become more conversative when nConds are considered. For example, Malaysian and Nepalese were positively significant for nWords, but become non-significant and showed negative trends for nConds. Even more interesting, the tests on the Simpson's diversity suggest that none of sub-ethnic groups show more diversity than the reference group. While many groups do not exhibit significant difference compared to the reference group, there are groups showing significantly lower diversity including Bangladeshi, Burmese, Filipino, Hmong, Korean, Malaysian, Nepalese, and Vietnamese. 
\begin{itemize}
    \item The South Asian group (including Indian, Pakistani, and Sri Lankan) consistently have higher values for nWords and nConds, indicating potential easier access to their information due to population size, familiarity with English, and cultural proximity among these ethnic groups. Their overall diversity measured in Simpson's diversity does not show significant difference with the reference group.  
    \item The Southeast Asian group (including Burmese, Laotian, and Vietnamese) consistently show lower values for all tests (the number of words, the number of conditions, and Simpson's diversity), reflecting less diversity in linguistic and health-conditional metrics. 
    \item While the East Asian group (including China, Japanese, and Korean) were most similar to Asian American reference in earlier results with RQ1,  disaggregated information for them seem insufficient, possibly due to barriers to language (i.e., English). 
\end{itemize}

\subsection{RQ3: How accurate is health information represented in LLM for various ethnic minority groups}
In this section, we investigate the accuracy and completeness of the responses generated for each subgroup. Specifically, we aim to identify instances of false information and evaluate whether any critical health-related details are missing. This analysis involves qualitative methods, including manual verification and review of the extracted conditions. By examining the reliability and comprehensiveness of the outputs, we seek to understand the extent to which LLMs faithfully represent health information for each subgroup and highlight areas where improvement is needed.

\subsubsection{Methods}\label{sec:RQ3_method}
In the following, we describe the methods used to extract health conditions from the responses and the approach taken to verify the validity of the extracted information. The extraction process involves identifying and categorizing health-related terms within the responses, while the verification process combines AI-assisted ``searches'' and manual review to assess the accuracy and completeness of the identified conditions.

\paragraph{Health condition extraction}

To extract health conditions from the collected responses, simple string manipulation using regular expressions is employed. Per each group, we count all conditions and sort them in a decreasing order in terms of occurrence frequencies. In an effort to reduce redundancy in terms of number of conditions (e.g., `healthcare disparities', `health care disparities', `disparities in healthcare'), but also maintain the distinction between similar terms that represent different conditions (e.g., `limited preventive care' and `preventive care'), three human workers manually go through all extracted conditions and develop a mapping of conditions. Once the manual mapping is designed, a script is used to clean the extracted health conditions by clubbing conditions together based on the mapping. Following this, conditions with occurrence counts smaller than 5 are dropped and not used in the further analysis.

\paragraph{Verification of extracted information} To check the validity of the extracted information, we take an extra round of verification by searching and confirming if there is a reliable source of information on the Web. That is, we seek to verify each condition in the processed dataset with a question, ``is a particular condition common among a particular race group?''. To this end, we take a two-step procedure: AI-assisted search of information source, followed by human verification. For the AI-assisted search, we utilize Google's Gemini API~\cite{gemini_api} with Google search grounding, where grounding is a technique to reduce hallucination in large language models by providing reliable data sources for the LLM to work with. The exact prompt given is 
\begin{quote}
\textit{``Is the [condition] more common among the [race group]?''}
\end{quote}
The Gemini API is instructed to provide its response in a format containing an ``Answer'' and a ``Reason'', where the later information is utilized by human workers to verify the information source. To ensure that the answer is based strictly on the Google search results, a parameter controlling the creativity Gemini is allowed to have is set to 0.

\subsubsection{Analysis}
We perform an analysis on the results obtained from the above process, which includes extraction and manual evaluation of health conditions. We report notable conditions that are validated as accurate, as well as highlight significant conditions that are missing from the responses.

\begin{table}[h]
\centering
\renewcommand{\arraystretch}{1.1}
\resizebox{\columnwidth}{!}{%
\begin{tabular}{c|c|c|c}
\toprule
\textbf{Group} & \textbf{\#. of Cond.} & \textbf{\#. of Unique.} & \textbf{\#. of Overlap.} \\
\midrule
Asian American     & 25 & 0 & - \\
\hline
Bangladeshi American    & 29 & 0 & 21 \\
Bhutanese American    & 32 & 0 & 19 \\
Burmese American & 40 & 0 & 20 \\
Cambodian American    & 32 & 1 & 16 \\
Chinese American    & 29 & 1 & 21\\
Filipino American    & 26 & 0 & 20\\
Hmong American & 40 & 3 & 16\\
Indian American    & 34 & 3 & 19\\
Indonesian American    & 32 & 0 & 21\\
Japanese American    & 23 & 0 & 17\\
Korean American & 25 & 1 & 19\\
Laotian American    & 33 & 1 & 20\\
Malaysian American    & 27 & 0 & 18\\
Mongolian American    & 31 & 0 & 19\\
Nepalese American & 39 & 1 & 21\\
Okinawan American    & 28 & 3 & 19\\
Pakistani American    & 29 & 0 & 20\\
Sri Lankan American    & 31 & 0 & 20\\
Taiwanese American & 26 & 0 & 19\\
Thai American    & 31 & 0 & 21\\
Vietnamese American    & 26 & 0 & 29\\
\bottomrule
\end{tabular}
}
\vspace{1mm}
\caption{The number of unique health conditions in each group (the second column), the number of those that are unique in each group (the third column), and the number of those overlapped with the reference group (`Asian', the fourth column).}
\label{tab:health_cond_count_refined}
\end{table}

\begin{table*}[t]
\centering
\begin{tabular}{p{0.11\textwidth}|p{0.82\textwidth}}
\toprule
\textbf{Group} & \textbf{Health Issues} \\
\hline
Asian & mental health issues, hepatitis b, osteoporosis, cancer, diabetes, heart diseases, cardiovascular disease, obesity, vision health, access to healthcare, tuberculosis, acculturation challenges, chronic diseases, communication barriers, liver health, tobacco use, vitamin d deficiency, respiratory illnesses, hypertension, hepatitis c, substance abuse, stroke,  cardiovascular issues, thyroid conditions, chronic kidney disease \\

\midrule
Bangladeshi & reproductive health, infectious diseases,  malnutrition, iron deficiency anemia,  dental health,  oral health issues, vitamin deficiencies, digestive health issues \\

\midrule
Bhutanese & infectious diseases, reproductive health, diet and nutrition, dental health,  chronic respiratory diseases, physical activity,  oral health issues,  chronic pain, malnutrition, trauma and PTSD, refugee trauma,  environmental health, chronic stress \\

\midrule
Burmese &  infectious diseases, diet and nutrition, malnutrition, refugee health concerns, reproductive health, mental health disparities, mental health stigma, environmental health, refugee trauma, trauma and PTSD, oral health issues, limited health literacy, chronic disease management, physical activity, dental health, preventive care,  genetic predispositions, food insecurity, chronic pain, traditional medicine \\

\midrule
Cambodian & diet and nutrition, oral health issues, mental health disparities, malnutrition, dental health, limited health literacy, reproductive health, infectious diseases, domestic violence, trauma and PTSD, \textbf{intergenerational trauma}, environmental health, chronic pain, \textbf{insufficient physical activity}, refugee health concerns, chronic respiratory diseases \\

\midrule
Chinese &  mental health stigma, mental health disparities,  traditional medicine, environmental health, infectious diseases, \textbf{hepatocellular carcinoma}, diet and nutrition \\

\midrule
Filipino &  infectious diseases, asthma, reproductive health, diet and nutrition, health disparities \\

\midrule
Hmong & reproductive health, traditional medicine, health disparities, limited health literacy, diet and nutrition, environmental health, cultural practices, infectious diseases, refugee trauma, chronic disease management,  preventive care, physical activity, trauma and PTSD, food insecurity, \textbf{immunization and vaccination}, chronic disease prevention, refugee health concerns, malnutrition, healthcare disparities, chronic disease prevalence, chronic pain, genetic predispositions, \textbf{healthy lifestyle choices}, \textbf{maternal health} \\

\midrule
Indian &  digestive health issues, gastrointestinal disorders, reproductive health, \textbf{metabolic syndrome}, eye health,  malnutrition, \textbf{gallbladder disease},  iron deficiency anemia, infectious diseases, sleep disorders,  asthma, \textbf{high cholesterol}, thalassemia, healthy weight, oral health issues \\
\midrule
Indonesian &  infectious diseases, malnutrition,  reproductive health,  diet and nutrition, dengue fever, environmental health, dental health, chronic disease prevalence, respiratory infections, physical activity, chronic pain \\
\bottomrule
\end{tabular}
\vspace{1mm}
\caption{Lists of health conditions of each group. Conditions in each list are presented in decreasing order of frequency. Conditions unique to each group are annotated in \textbf{bold}; those already listed under the reference group (``Asian American'') are not repeated. [Part 1]}
\label{tab:health_issues_r1}
\end{table*}

\paragraph{Accurately Represented Conditions} Some well-known health conditions appear in nearly all demographic groups, suggesting that the LLM responses are very well reflecting common health concerns: Mental health issues, cardiovascular diseases (including hypertension, heart diseases, and strokes), diabetes, hepatitis B,  obesity, cancer, respiratory illnesses, and access to healthcare challenges. It is also well-reflected that the liver health is a bigger concern in the Asian population~\cite{pham2018striking}. Also, the LLM responses reflect the unique group specific issues: (1) PSTD and Trauma emphsized in Bhutanese~\cite{frounfelker2023mental}, Burmese~\cite{lim2013trauma}, Cambodian~\cite{marshall2006rates}, Hmong~\cite{collier2012hmong}, and Laotian~\cite{gordon2011trauma} groups , (2) ``Aging and longevity'' and ``Rheumatoid arthritis and Alzheimer's disease'' (which are common in the elderly population) in the Okinawan group, and (3) Metabolic syndrome in the South Asian groups, particularly, gallbladder disease \cite{kapoor2003gallbladder} (or gallbladder cancer) is considered more common in people of Indian origin (often time called, `Indian' cancer). 

\paragraph{Uncaptured health conditions} While the common health conditions are properly captured, certain critical details are often poorly captured or missing. Table~\ref{tab:health_cond_count_refined} reports the number of extracted health conditions, the number of conditions that are unique to each group, and the number of conditions that are overlapping with the reference group, Asian American. The average ratio of the number of overlapping conditions to the number of conditions in each group (averaged across the group) is 0.63, suggesting that general terms are used to describe more than half of conditions in each group. Also, very small portion of the unique conditions in each group suggest that there are many overlapping terms in between groups. 

\begin{table*}[t]
\centering
\begin{tabular}{p{0.12\textwidth}|p{0.82\textwidth}}
\toprule
\textbf{Group} & \textbf{Health Issues} \\
\midrule
Japanese &  eye health, bone health, sleep disorders, allergies and asthma, aging population, infectious diseases \\

\midrule
Korean &  mental health stigma,  mental health disparities,  smoking, diet and nutrition, eye health, \textbf{injuries and accidents} \\

\midrule
Laotian & limited health literacy, chronic pain, reproductive health, smoking, malnutrition, health disparities, physical activity, infectious diseases, \textbf{limited preventive care}, oral health issues, dental health, trauma and PTSD\\

\midrule
Malaysian & infectious diseases, dengue fever, eye health, dental health, diet and nutrition, reproductive health, respiratory infections, malnutrition, oral health issues \\

\midrule
Mongolian & infectious diseases, mental health disparities, reproductive health, diet and nutrition, malnutrition, mental health stigma, physical activity, bone health, environmental health,  genetic predispositions, digestive health issues, cultural practices \\

\midrule
Nepalese & reproductive health, diet and nutrition, mental health stigma, physical activity, environmental health, malnutrition, mental health disparities, \textbf{mental health awareness},  preventive care, dental health, oral health issues, genetic predispositions, bone health, chronic disease management, chronic disease prevention, iron deficiency anemia, healthcare disparities \\
\midrule
Okinawan &  bone health,  infectious diseases, \textbf{aging and longevity},  eye health, dental health, aging population,  \textbf{rheumatoid arthritis}, \textbf{alzheimer's disease}, mental health disparities \\

\midrule
Pakistani & reproductive health, infectious diseases,  iron deficiency anemia,  thalassemia, oral health issues, smoking,  dental health,  gastrointestinal disorders, malnutrition \\

\midrule
SriLankan &  infectious diseases, digestive health issues, reproductive health,  malnutrition, oral health issues,  dental health, vitamin deficiencies, bone health, gastrointestinal disorders, iron deficiency anemia, genetic predispositions \\

\midrule
Taiwanese &  eye health, bone health, sleep disorders, diet and nutrition, infectious diseases, chronic disease prevention, physical activity \\

\midrule
Thai &  bone health, reproductive health, oral health issues, chronic pain, eye health, infectious diseases,  dental health, domestic violence, physical activity \\

\midrule
Vietnamese &  smoking, diet and nutrition,  chronic pain,  infectious diseases, dental health, oral health issues \\
\bottomrule
\end{tabular}
\vspace{1mm}
\caption{Lists of health conditions of each group. Conditions in each list are presented in an order of decreasing frequencies. The names of conditions that appear uniquely in each group are annotated in \textbf{bold} fonts and the name of conditions that appear in the reference group (``Asian American'') are not repeated in other groups. [Part 2]}
\label{tab:health_issues_r2}
\end{table*}

Motivated by the result, we investigate if there are any conditions that are known to be more prevalent in medical domains, but not captured in the LLM response. For this purpose, we refer to a report on cancer health disparities in racial/ethnic minorities in the US~\cite{zavala2021cancer}. Examples of such missing conditions include:
\begin{itemize}
    \item Cervical Cancer in Hmong Women: The cervical cancer rate among Hmong women in California is strikingly high, at 36.6 per 100,000. This rate is more than three times the aggregated rate for Asian Pacific Islanders (11.8 per 100,000) and more than four times the rate for non-Hispanic White women in California (8.0 per 100,000). These disparities highlight significant health inequities within the Hmong community~\cite{fang2010factors}.
    \item Cervical Cancer in Vietnamese and Cambodian Women: Vietnamese and Cambodian women experience disproportionately high rates of cervical cancer compared to other groups~\cite{torre2016cancer}, underscoring the need for more targeted health interventions and awareness campaigns. However, this information is not consistently or clearly presented in LLM responses.
    \item Breast Cancer in Specific Asian Subgroups: Korean, Filipina, Vietnamese, and Chinese women are at a significantly increased risk of being diagnosed with HER2-positive breast cancer subtypes compared to non-Hispanic White women~\cite{telli2011asian}. Younger US-born Chinese and Filipina women (under 55 years old) have higher rates of HER2-positive breast cancer than White women of the same age. Moreover, the rates of HER2-positive breast cancer have been increasing over time, with particularly sharp increases observed in foreign-born Korean women and US-born Filipina women, at rates as high as 4\% per year.
    \item Gastric Cancer in Korean American and Japanese American Populations: Korean Americans and Japanese Americans are at a significantly higher risk of developing gastric cancer compared to other ethnic groups~\cite{lee2017stomach}, indicating the need for focused health screening and prevention strategies within these communities.
    \item Liver Cancer in Southeast Asian Populations: Liver cancer, particularly hepatocellular carcinoma (HCC), is more prevalent among Asians in California compared to non-Hispanic Whites, African Americans, and Hispanics/Latinos. Within Asian populations, the incidence of HCC is especially high among Southeast Asians, including Laotians, Vietnamese, and Cambodians, who have rates eight to nine times higher than other Asian groups~\cite{pham2018striking}. This critical information is often found in sources written in Chinese, making it less accessible to English-speaking audiences.
\end{itemize}

The qualitative analysis of the collected responses highlights both strengths and gaps in representing subgroup-specific health information. Common health conditions like mental health issues, cardiovascular diseases, diabetes, and cancer are accurately captured across most groups. Additionally, unique concerns, such as PTSD and trauma in Southeast Asian subgroups, aging-related conditions in Okinawans, and metabolic syndrome in South Asians, are well-represented. However, significant omissions remain, including critical disparities like high rates of cervical cancer among Hmong, Vietnamese, and Cambodian women, and elevated risks of HER2-positive breast cancer, gastric cancer, and liver cancer in specific subgroups. These gaps suggest that the responses often rely on general terms, with insufficient attention to subgroup-specific conditions. Improving data disaggregation and integrating more authoritative medical sources is essential to enhance the accuracy and completeness of LLM outputs.

\section{Related Work}
\paragraph{Stereotyping and bias in LLMs} To the best of our knowledge, no prior work has specifically investigated the practice of using disaggregated data with LLMs. However, recent studies have thoroughly explored the issue of stereotyping in LLMs, highlighting the persistence and complexity of biases in these systems. In \cite{cheng2023marked}, a prompt-based method, Marked Personas, is introduced to to measure stereotypes by comparing LLM-generated personas of intersectional demographic groups with unmarked ones, finding that these portrayals are more stereotypical than those written by humans. In \cite{ma2023intersectional}, the work of \cite{cheng2023marked} is further extended to investigate stereotypes in LLMs, presenting a dataset of intersectional biases generated with ChatGPT and validated by human reviewers, emphasizing the persistent need for improvements in addressing stereotypes. In \cite{ghosh2023chatgpt}, gender bias in machine translation is studied. In  \cite{kirk2021bias,kotek2023gender}, the presence of intersectional occupational biases in LLMs is investigated, while in \cite{kotek2023gender}, a method for debiasing gender bias in job advertisement generated via LLMs is studied. In \cite{mei2023bias}, studies on bias in LLMs are further extended to analyze biases against finer categorizations of intersectional or stigmatized groups. In early studies, there are works investigating intersectional biases in LLMs (e.g., GPT-2)  \cite{tan2019assessing}, creating a dataset consisting of 600 intersectional bias descriptors \cite{smith2022m}, and measuring toxicity and negative sentiment in biases \cite{esiobu2023robbie}. As a method for mitigating biases, a fine-tuning method using an interventioned dataset is proposed in \cite{thakur2023language}. Data disaggregation is a widely used practice in the medical domain, as shown in the references utilized in this study. However, its application in other fields remains limited. While it has been emphasized in social studies, as proposed in \cite{wu2024not}, there is a lack of extensive research utilizing this approach.

\section{Discussion \& Conclusion}
In this study, we utilized various statistical methods to address our research questions and assess the capabilities of LLMs in representing health-related information across Asian American subgroups. Our findings reveal both strengths and limitations in the ability of LLMs to capture subgroup-specific health conditions, highlighting the importance of disaggregation for equitable data representation.

Chinese, Taiwanese, and Korean subgroups emerged as the most representative of the broader Asian American category, as evidenced by their low distances in embedding spaces. However, even within these well-represented groups, certain critical health conditions—such as breast cancer in Chinese women and gastric cancer in Koreans—are underrepresented or overly generalized. This systemic tendency to rely on generic representations underscores a broader limitation in the ability of LLMs to address subgroup-specific health disparities comprehensively.

While Chinese, Taiwanese, and Korean subgroups closely align with the broader Asian American reference category, this trend is not uniform across all East Asian populations. Japanese, Okinawan, and Mongolian groups exhibit weaker representation, characterized by greater distances from the reference category, fewer identified health conditions, and lower response diversity. For example, Japanese and Okinawan health conditions primarily center on age-related topics, such as “aging and longevity” and “rheumatoid arthritis and Alzheimer's disease.” Although disaggregated prompting successfully identifies these significant issues, other critical conditions, like gastric cancer prevalent among Japanese populations, are notably absent. These gaps emphasize the need for more targeted prompts and improved data equity to capture the full scope of subgroup-specific health challenges.

Among South Asian groups, significant variability was observed. Subgroups such as Indian, Pakistani, Sri Lankan, and Bangladeshi demonstrated more diverse health conditions, with robust representation of issues like metabolic syndrome, gallbladder diseases, vitamin deficiency, bone-related conditions, and mental health challenges. Conversely, Bhutanese and Nepalese subgroups displayed limited representation, with fewer identified health conditions and reduced diversity in responses.

A similar pattern emerged among Southeast Asian subgroups. Except for Filipinos, these groups generally exhibited lower representation in terms of both the number of identified conditions and response diversity relative to Asian Americans as a whole. While some specific health concerns, such as PTSD, trauma, and oral health issues, were captured, significant gaps remain. For instance, cervical cancer, which is highly prevalent among Hmong women, and liver cancer, particularly affecting Vietnamese populations, are often missing. Filipinos stood out with higher representation and greater diversity in responses; however, important conditions like breast cancer was still absent from their health profiles.

In summary, while disaggregated prompting demonstrates promise in capturing some subgroup-specific health disparities, there are significant limitations in both accuracy and comprehensiveness. These findings highlight the need for more refined prompt design, better integration of authoritative data sources, and an ongoing commitment to data equity to ensure that the health challenges of diverse populations are accurately and equitably represented in LLM outputs.
\UB{\paragraph{Limitations and Future Work} This study exclusively uses ChatGPT-3.5. While this allows for focused evaluation, it limits the generalizability of our findings to other LLMs. Still, our results highlight information inequality in LLMs, and extending this analysis to other models is a key direction for future research. Additionally, our reliance on Census-based ethnic categorizations limits the ability to capture intra-group diversity. Notably, distinctions within multi-ethnic groups—such as regional or linguistic differences in Indian or Malaysian populations—remain unaddressed.}

\clearpage

\section{Acknowledgment}

UBM, BJ, KL, and KHK acknowledge the support from the U.S. National Science Foundation under grant CNS2210137. KHK acknowledges the partial support from the Seed Grant Program at Arizona State University's Institute for Social Science Research.

\section{Ethical Considerations Statement}
This study involves the use of disaggregated demographic data and the analysis of outputs from large language models (LLMs). The analysis presented in this study is grounded in medical information that we did our best to cross-reference with peer-reviewed journal and conference publications. Based on this effort, we do not foresee any potential ethical violations arising from this work.

\section{Adverse Impact Statement}
A potential unintended impact of this work lies in the possibility of misunderstanding the findings. While the study aims to provide nuanced insights into representation equity using disaggregated demographic data, there is a risk that the results could be misinterpreted or taken out of context. For instance, entities might use the findings to support narrow, biased narratives about specific groups rather than advocating for equity.

\bibliography{aaai25}

\begin{thebibliography}{47}
\providecommand{\natexlab}[1]{#1}

\bibitem[{AI(2023{\natexlab{a}})}]{gemini_api}
AI, G. 2023{\natexlab{a}}.
\newblock {Gemini API}.
\newblock Google Cloud Vertex AI.
\newblock Accessed on 2025-01-16.

\bibitem[{AI(2023{\natexlab{b}})}]{perplexity2023}
AI, P. 2023{\natexlab{b}}.
\newblock Perplexity AI: A Large Language Model.

\bibitem[{Anthropic(2023)}]{anthropic2023claude}
Anthropic. 2023.
\newblock Claude: A Family of Language Models.

\bibitem[{Bouma(2009)}]{Bouma2009NormalizedPMI}
Bouma, G. 2009.
\newblock Normalized (pointwise) mutual information in collocation extraction.
\newblock In \emph{Proceedings of the GSCL}, volume~30, 31--40.

\bibitem[{Brown et~al.(2020)Brown, Mann, Ryder, Subbiah, Kaplan, Dhariwal, Neelakantan, Shyam, Sastry, Askell, and et~al.}]{brown2020language}
Brown, T.~B.; Mann, B.; Ryder, N.; Subbiah, M.; Kaplan, J.; Dhariwal, P.; Neelakantan, A.; Shyam, P.; Sastry, G.; Askell, A.; and et~al. 2020.
\newblock Language Models are Few-Shot Learners.
\newblock \emph{arXiv preprint arXiv:2005.14165}.

\bibitem[{{Centers for Disease Control and Prevention}(2020)}]{CDC2020}
{Centers for Disease Control and Prevention}. 2020.
\newblock {WISQARS (Web-based Injury Statistics Query and Reporting System)}.
\newblock \url{https://www.cdc.gov/injury/wisqars/index.html}.
\newblock {Retrieved from \url{https://www.cdc.gov/injury/wisqars/index.html}}.

\bibitem[{Cheng, Durmus, and Jurafsky(2023)}]{cheng2023marked}
Cheng, M.; Durmus, E.; and Jurafsky, D. 2023.
\newblock Marked Personas: {U}sing Natural Language Prompts to Measure Stereotypes in Language Models.
\newblock In \emph{Proceedings of the 61st Annual Meeting of the Association for Computational Linguistics (Volume 1: Long Papers)}, 1504--1532.

\bibitem[{Collier, Munger, and Moua(2012)}]{collier2012hmong}
Collier, A.~F.; Munger, M.; and Moua, Y.~K. 2012.
\newblock Hmong mental health needs assessment: A community-based partnership in a small mid-western community.
\newblock \emph{American Journal of Community Psychology}, 49: 73--86.

\bibitem[{Dieng, Ruiz, and Blei(2020)}]{Dieng2020TopicModeling}
Dieng, A.~B.; Ruiz, F. J.~R.; and Blei, D.~M. 2020.
\newblock Topic modeling in embedding spaces.
\newblock \emph{Transactions of the Association for Computational Linguistics}, 8: 439--453.

\bibitem[{Esiobu et~al.()Esiobu, Tan, Hosseini, Ung, Zhang, Fernandes, Dwivedi-Yu, Presani, Williams, and Smith}]{esiobu2023robbie}
Esiobu, D.; Tan, X.; Hosseini, S.; Ung, M.; Zhang, Y.; Fernandes, J.; Dwivedi-Yu, J.; Presani, E.; Williams, A.; and Smith, E.~M. ????
\newblock ROBBIE: Robust Bias Evaluation of Large Generative Language Models.
\newblock In \emph{The 2023 Conference on Empirical Methods in Natural Language Processing}.

\bibitem[{Fang et~al.(2010)Fang, Lee, Stewart, Ly, and Chen~Jr}]{fang2010factors}
Fang, D.~M.; Lee, S.; Stewart, S.; Ly, M.~Y.; and Chen~Jr, M.~S. 2010.
\newblock Factors associated with {Pap} testing among {H}mong women.
\newblock \emph{Journal of health care for the poor and underserved}, 21(3): 839--850.

\bibitem[{Frounfelker et~al.(2023)Frounfelker, Mishra, Holmes, Gautam, and Betancourt}]{frounfelker2023mental}
Frounfelker, R.~L.; Mishra, T.; Holmes, K.~B.; Gautam, B.; and Betancourt, T.~S. 2023.
\newblock Mental health among older Bhutanese with a refugee life experience: A mixed-methods latent class analysis study.
\newblock \emph{American Journal of Orthopsychiatry}, 93(4): 304.

\bibitem[{Ghosh and Caliskan(2023)}]{ghosh2023chatgpt}
Ghosh, S.; and Caliskan, A. 2023.
\newblock Chatgpt perpetuates gender bias in machine translation and ignores non-gendered pronouns: {F}indings across bengali and five other low-resource languages.
\newblock In \emph{Proceedings of the 2023 AAAI/ACM Conference on AI, Ethics, and Society}, 901--912.

\bibitem[{Gordon(2011)}]{gordon2011trauma}
Gordon, D.~M. 2011.
\newblock Trauma and second language learning among Laotian refugees.
\newblock \emph{Journal of Southeast Asian American Education \& Advancement}, 6(1): 1--17.

\bibitem[{Grootendorst(2022)}]{grootendorst2022bertopic}
Grootendorst, M. 2022.
\newblock {BERTopic}: {N}eural topic modeling with a class-based {TF-IDF} procedure.
\newblock \emph{arXiv:2203.05794}.

\bibitem[{Kapoor and McMichael(2003)}]{kapoor2003gallbladder}
Kapoor, V.; and McMichael, A. 2003.
\newblock Gallbladder cancer: {A}n ‘{I}ndian’disease.
\newblock \emph{Natl Med J India}, 16(4): 209--213.

\bibitem[{Kirk et~al.(2021)Kirk, Jun, Volpin, Iqbal, Benussi, Dreyer, Shtedritski, and Asano}]{kirk2021bias}
Kirk, H.~R.; Jun, Y.; Volpin, F.; Iqbal, H.; Benussi, E.; Dreyer, F.; Shtedritski, A.; and Asano, Y. 2021.
\newblock Bias out-of-the-box: {A}n empirical analysis of intersectional occupational biases in popular generative language models.
\newblock \emph{Advances in neural information processing systems}, 34: 2611--2624.

\bibitem[{Kotek, Dockum, and Sun(2023)}]{kotek2023gender}
Kotek, H.; Dockum, R.; and Sun, D. 2023.
\newblock Gender bias and stereotypes in large language models.
\newblock In \emph{Proceedings of the ACM collective intelligence conference}, 12--24.

\bibitem[{Lee et~al.(2017)Lee, Liu, Zhang, Stern, Barzi, Hwang, Kim, Hamilton, Wu, and Deapen}]{lee2017stomach}
Lee, E.; Liu, L.; Zhang, J.; Stern, M.~C.; Barzi, A.; Hwang, A.; Kim, A.~E.; Hamilton, A.~S.; Wu, A.~H.; and Deapen, D. 2017.
\newblock {Stomach cancer disparity among Korean Americans by tumor characteristics: comparison with non-Hispanic whites, Japanese Americans, South Koreans, and Japanese}.
\newblock \emph{Cancer Epidemiology, Biomarkers \& Prevention}, 26(4): 587--596.

\bibitem[{Lim et~al.(2013)Lim, Stock, Shwe~Oo, and Jutte}]{lim2013trauma}
Lim, A.~G.; Stock, L.; Shwe~Oo, E.~K.; and Jutte, D.~P. 2013.
\newblock Trauma and mental health of medics in eastern Myanmar’s conflict zones: a cross-sectional and mixed methods investigation.
\newblock \emph{Conflict and health}, 7: 1--13.

\bibitem[{Ma et~al.(2023)Ma, Chiang, Wu, Wang, and Vosoughi}]{ma2023intersectional}
Ma, W.; Chiang, B.; Wu, T.; Wang, L.; and Vosoughi, S. 2023.
\newblock Intersectional stereotypes in large language models: {D}ataset and analysis.
\newblock In \emph{Findings of the Association for Computational Linguistics: EMNLP 2023}, 8589--8597.

\bibitem[{Marshall et~al.(2006)Marshall, Berthold, Schell, Elliott, Chun, and Hambarsoomians}]{marshall2006rates}
Marshall, G.~N.; Berthold, S.~M.; Schell, T.~L.; Elliott, M.~N.; Chun, C.-A.; and Hambarsoomians, K. 2006.
\newblock Rates and correlates of seeking mental health services among Cambodian refugees.
\newblock \emph{American Journal of Public Health}, 96(10): 1829--1835.

\bibitem[{McInnes, Healy, and Astels(2017)}]{McInnes2017HDBSCAN}
McInnes, L.; Healy, J.; and Astels, S. 2017.
\newblock {HDBSCAN: H}ierarchical density based clustering.
\newblock \emph{Journal of Open Source Software}, 2(11): 205.

\bibitem[{McInnes et~al.(2018)McInnes, Healy, Saul, and Grossberger}]{McInnes2018UMAP}
McInnes, L.; Healy, J.; Saul, N.; and Grossberger, L. 2018.
\newblock {UMAP: U}niform Manifold Approximation and Projection.
\newblock \emph{The Journal of Open Source Software}, 3(29): 861.

\bibitem[{Mei, Fereidooni, and Caliskan(2023)}]{mei2023bias}
Mei, K.; Fereidooni, S.; and Caliskan, A. 2023.
\newblock Bias against 93 stigmatized groups in masked language models and downstream sentiment classification tasks.
\newblock In \emph{Proceedings of the 2023 ACM Conference on Fairness, Accountability, and Transparency}, 1699--1710.

\bibitem[{Movva et~al.(2023)Movva, Shanmugam, Hou, Pathak, Guttag, Garg, and Pierson}]{movva2023coarse}
Movva, R.; Shanmugam, D.; Hou, K.; Pathak, P.; Guttag, J.; Garg, N.; and Pierson, E. 2023.
\newblock Coarse race data conceals disparities in clinical risk score performance.
\newblock In \emph{Machine Learning for Healthcare Conference}, 443--472. PMLR.

\bibitem[{{National Cancer Institute}(2020)}]{NCI2020}
{National Cancer Institute}. 2020.
\newblock {SEER Cancer Statistics Review, 1975-2017}.
\newblock \url{https://seer.cancer.gov/}.
\newblock {Retrieved from \url{https://seer.cancer.gov/}}.

\bibitem[{{National Center for Education Statistics}(2021)}]{NCES2021}
{National Center for Education Statistics}. 2021.
\newblock {The condition of education 2021}.
\newblock \url{https://nces.ed.gov/programs/coe/}.
\newblock {Retrieved from \url{https://nces.ed.gov/programs/coe/}}.

\bibitem[{OpenAI(2024)}]{openai_api}
OpenAI. 2024.
\newblock OpenAI API.

\bibitem[{{Pew Research Center}(2019)}]{Pew2019}
{Pew Research Center}. 2019.
\newblock {Key facts about Asian Americans, a diverse and growing population}.
\newblock \url{https://www.pewresearch.org/}.
\newblock {Retrieved from \url{https://www.pewresearch.org/}}.

\bibitem[{{Pew Research Center}(2021)}]{Pew2021}
{Pew Research Center}. 2021.
\newblock {Income inequality in the U.S. is rising most rapidly among Asians}.
\newblock \url{https://www.pewresearch.org/}.
\newblock {Retrieved from \url{https://www.pewresearch.org/}}.

\bibitem[{Pham et~al.(2018)Pham, Fong, Zhang, and Liu}]{pham2018striking}
Pham, C.; Fong, T.-L.; Zhang, J.; and Liu, L. 2018.
\newblock Striking racial/ethnic disparities in liver cancer incidence rates and temporal trends in {C}alifornia, 1988--2012.
\newblock \emph{JNCI: Journal of the National Cancer Institute}, 110(11): 1259--1269.

\bibitem[{Reimers and Gurevych(2019)}]{Reimers2019SentenceBERT}
Reimers, N.; and Gurevych, I. 2019.
\newblock Sentence-{BERT: S}entence Embeddings using {S}iamese {BERT}-Networks.
\newblock In \emph{Proceedings of the 2019 Conference on Empirical Methods in Natural Language Processing}. Association for Computational Linguistics.

\bibitem[{Sasa and Yellow~Horse(2022)}]{sasa2022just}
Sasa, S.~M.; and Yellow~Horse, A.~J. 2022.
\newblock Just data representation for Native Hawaiians and Pacific Islanders: A critical review of systemic Indigenous erasure in census and recommendations for psychologists.
\newblock \emph{American journal of community psychology}, 69(3-4): 343--354.

\bibitem[{Shimkhada, Scheitler, and Ponce(2021)}]{shimkhada2021capturing}
Shimkhada, R.; Scheitler, A.; and Ponce, N.~A. 2021.
\newblock Capturing racial/ethnic diversity in population-based surveys: data disaggregation of health data for Asian American, Native Hawaiian, and Pacific Islanders (AANHPIs).
\newblock \emph{Population Research and Policy Review}, 40(1): 81--102.

\bibitem[{Smith et~al.(2022)Smith, Hall, Kambadur, Presani, and Williams}]{smith2022m}
Smith, E.~M.; Hall, M.; Kambadur, M.; Presani, E.; and Williams, A. 2022.
\newblock “I’m sorry to hear that”: {F}inding New Biases in Language Models with a Holistic Descriptor Dataset.
\newblock In \emph{Proceedings of the 2022 Conference on Empirical Methods in Natural Language Processing}, 9180--9211.

\bibitem[{Stonier et~al.(2023)Stonier, Woodman, Alshammari, Cummings, Dad, Garg, Busetto, Hsiao, Hudson, Singh et~al.}]{stonier2023data}
Stonier, J.; Woodman, L.; Alshammari, M.; Cummings, R.; Dad, N.; Garg, A.; Busetto, A.~G.; Hsiao, K.; Hudson, M.; Singh, P.~J.; et~al. 2023.
\newblock Data Equity: Foundational Concepts for Generative AI.
\newblock \emph{arXiv preprint arXiv:2311.10741}.

\bibitem[{Tan and Celis(2019)}]{tan2019assessing}
Tan, Y.~C.; and Celis, L.~E. 2019.
\newblock Assessing social and intersectional biases in contextualized word representations.
\newblock \emph{Advances in neural information processing systems}, 32.

\bibitem[{Telli et~al.(2011)Telli, Chang, Kurian, Keegan, McClure, Lichtensztajn, Ford, and Gomez}]{telli2011asian}
Telli, M.~L.; Chang, E.~T.; Kurian, A.~W.; Keegan, T.~H.; McClure, L.~A.; Lichtensztajn, D.; Ford, J.~M.; and Gomez, S.~L. 2011.
\newblock {A}sian ethnicity and breast cancer subtypes: {A} study from the {C}alifornia Cancer Registry.
\newblock \emph{Breast cancer research and treatment}, 127: 471--478.

\bibitem[{Thakur et~al.(2023)Thakur, Jain, Vaddamanu, Liang, and Morency}]{thakur2023language}
Thakur, H.; Jain, A.; Vaddamanu, P.; Liang, P.~P.; and Morency, L.-P. 2023.
\newblock Language Models Get a Gender Makeover: {M}itigating Gender Bias with Few-Shot Data Interventions.
\newblock In \emph{The 61st Annual Meeting Of The Association For Computational Linguistics}.

\bibitem[{{The White House}(2022)}]{WhiteHouse2022}
{The White House}. 2022.
\newblock {Vision for Equitable Data}.
\newblock \url{https://www.whitehouse.gov/wp-content/uploads/2022/04/eo13985-vision-for-equitable-data.pdf}.
\newblock {Retrieved from \url{https://www.whitehouse.gov/wp-content/uploads/2022/04/eo13985-vision-for-equitable-data.pdf}}.

\bibitem[{the~{W}hite {H}ouse(2023)}]{whitehouse2023National}
the~{W}hite {H}ouse. 2023.
\newblock \emph{National Strategy to Advance Equity, Justice, and Opportunity for Asian American, Native Hawaiian, and Pacific Islander (AA and NHPI) Communities}.

\bibitem[{Torre et~al.(2016)Torre, Sauer, Chen~Jr, Kagawa-Singer, Jemal, and Siegel}]{torre2016cancer}
Torre, L.~A.; Sauer, A. M.~G.; Chen~Jr, M.~S.; Kagawa-Singer, M.; Jemal, A.; and Siegel, R.~L. 2016.
\newblock Cancer statistics for {Asian Americans, Native Hawaiians, and Pacific Islanders, 2016: Converging incidence in males and females}.
\newblock \emph{CA: a cancer journal for clinicians}, 66(3): 182--202.

\bibitem[{Tukey(1949)}]{tukey_hsd}
Tukey, J.~W. 1949.
\newblock Comparison of Means: A Powerful Method of Comparing Groups with Application to Experiments in Education and Psychology.
\newblock \emph{The Annals of Mathematical Statistics}, 20(2): 704--711.

\bibitem[{{United States Census Bureau}(2022)}]{census2022aanhpi}
{United States Census Bureau}. 2022.
\newblock The AANHPI Population is Diverse and Geographically Dispersed.

\bibitem[{Wu et~al.(2024)Wu, Lakhanpal, Li, Lee, Kim, Chae, and Kwon}]{wu2024not}
Wu, F.; Lakhanpal, S.; Li, Q.; Lee, K.; Kim, D.; Chae, H.; and Kwon, K.~H. 2024.
\newblock Not All Asians are the Same: {A} Disaggregated Approach to Identifying Anti-Asian Racism in Social Media.
\newblock In \emph{Proceedings of the ACM on Web Conference 2024}, 2615--2626.

\bibitem[{Zavala et~al.(2021)Zavala, Bracci, Carethers, Carvajal-Carmona, Coggins, Cruz-Correa, Davis, de~Smith, Dutil, Figueiredo et~al.}]{zavala2021cancer}
Zavala, V.~A.; Bracci, P.~M.; Carethers, J.~M.; Carvajal-Carmona, L.; Coggins, N.~B.; Cruz-Correa, M.~R.; Davis, M.; de~Smith, A.~J.; Dutil, J.; Figueiredo, J.~C.; et~al. 2021.
\newblock {Cancer health disparities in racial/ethnic minorities in the United States}.
\newblock \emph{British journal of cancer}, 124(2): 315--332.

\end{thebibliography}

\clearpage
\section{Research Methods}
\subsection{Regular expression used to extract conditions} \KL{The LLM outputs largely follows a numbered list pattern such that ``[Number]. [Condition]: [Description].'' The regular expression pattern is used to extract the "Condition" for our analyses. We provide the code snippet for regex here, ``\text{pattern = re.compile(r'\textbackslash 
d+\.\textbackslash 
s*([A-Za-z\textbackslash
 s]+)\textbackslash
 s*:')}''.}

\subsection{BERTopic details}\label{app:bertopic} We perform topic modeling using the BERTopic API \cite{grootendorst2022bertopic}. BERTopic clusters unstructured data by leveraging sentence embeddings, dimensionality reduction, and clustering. It assigns topics to clusters by identifying the most relevant words within each cluster using C-TF-IDF scores. For this, we utilize the Sentence-Transformer model ``all-MiniLM-L6-v2" \cite{Reimers2019SentenceBERT}, UMAP \cite{McInnes2018UMAP} for dimensionality reduction, and HDBSCAN \cite{McInnes2017HDBSCAN} for clustering. 
\KL{Moreover, in our preliminary study, we find that LDA typically results in lower performance in terms of topic coherence and topic diversity. Along with the degraded performance, LDA requires hyper-parameter tuning to find an appropriate value for the number of topics; on the other hand, BERTopic automatically determines the number of clusters in HDBSCAN. Due to this reasons, we choose BERTopic.
}

Each generation within each group contains substantial content and can cover multiple topics. To enhance performance, we tokenize the sentences and treat each sentence as a separate document, then run the BERTopic model for each group using 135 different combinations of hyperparameters. To determine the best model for each group, we use two metrics: topic coherence (TC) and topic diversity (TD). Statistical coherence methods assess how frequently words within a topic co-occur in documents, based on corpus statistics. The underlying idea is that words in a well-defined topic should frequently appear together in the same documents. Topic coherence is quantified using normalized pointwise mutual information (NPMI) \cite{Bouma2009NormalizedPMI}, which ranges from -1 to 1. A score of -1 indicates that the top $n$ words in a topic never co-occur, 0 indicates independence, and 1 represents complete co-occurrence of the top $n$ words. Topic diversity reflects how unique the topics are, calculated as the proportion of distinct words among the top 10 words in the top 10 topics \cite{Dieng2020TopicModeling}. The TD score ranges from 0 to 1, where 0 signifies redundant topics, and 1 represents a higher variety. A higher TD score indicates broader coverage of different aspects within the corpus.

\begin{figure}[h]
    \centering
    \includegraphics[width=0.9\columnwidth]{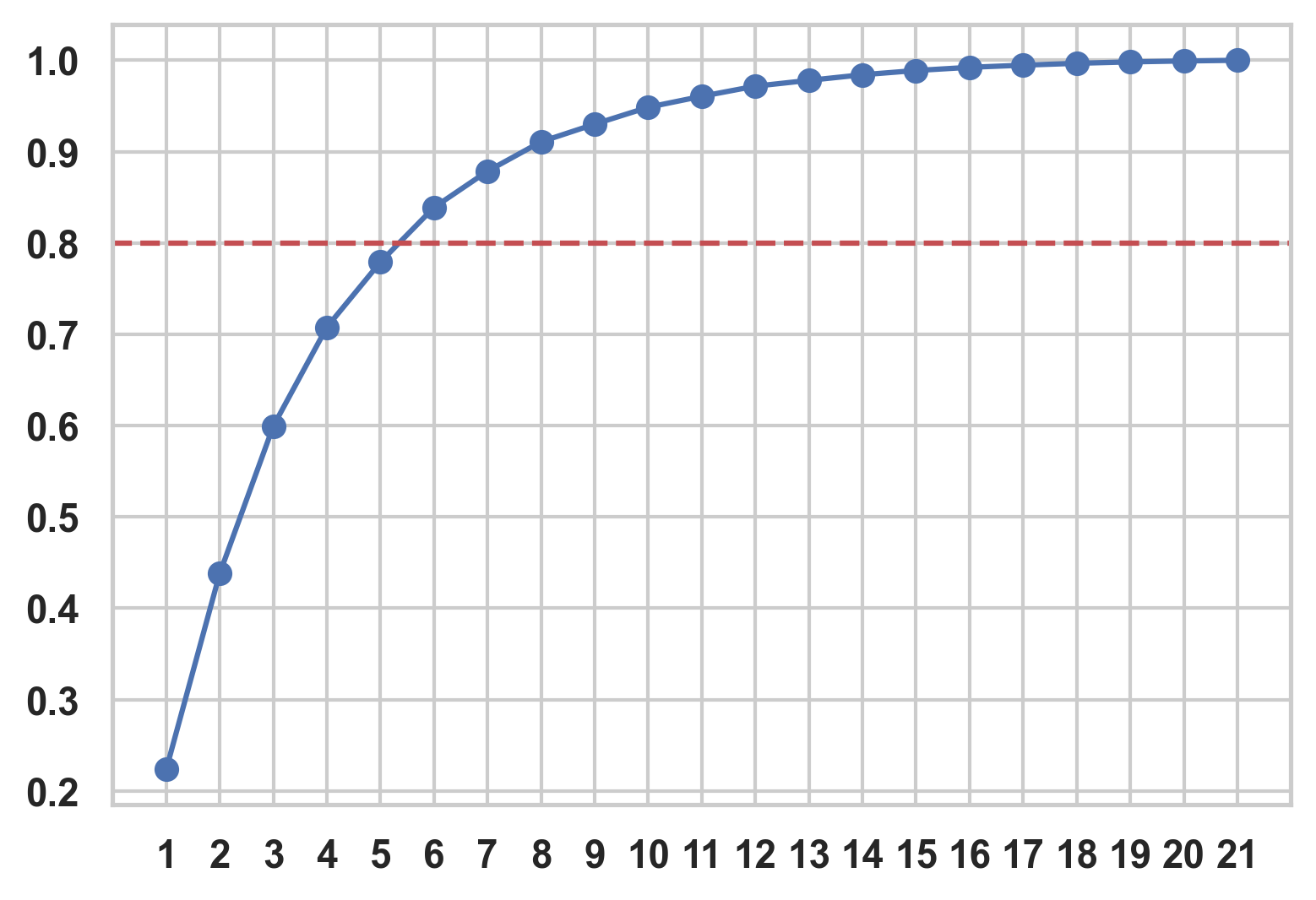}
    \caption{Cumulative variance explained by the correspondence analysis dimensions for the Asian American sub-ethnic demographics. The red dashed line indicates the 80\% threshold used to determine the optimal number of dimensions.}
    \label{fig:cumulative_variance_plot_R1}
\end{figure}

\subsection{Correspondence analysis details}\label{app:CA} To identify the clusters most strongly associated with each demographic group, we use a modified metric of ``relationship strength'' based on the CA results, where the relationship strength is essentially measured as an inner product between two embedding vectors: One vector associated with demography categorical variables and another vector associated with cluster categorical variables. After transforming the demographic groups and clusters into a lower-dimensional space, we determine the optimal number of dimensions that explain 80\% of the cumulative variance, based on the eigenvalues of the CA decomposition (see Figure~\ref{fig:cumulative_variance_plot_R1}). After computing the relationship strength for each demographic-cluster pair, we rank the clusters based on their strength. The top 15 clusters with the highest relationship strength for each demographic are identified as the closest clusters. 

\begin{table*}[t]
\centering
\begin{tabular}{p{0.1\textwidth}|p{0.85\textwidth}}
\toprule
\textbf{Group} & \textbf{Closest topic clusters} \\
\midrule
\textbf{Asian} & 
\textit{[31]} thyroid, disorders, hypothyroidism, digestive, bowel, irritable, syndrome, hyperthyroidism, women, intolerance \\
& \textit{[4]} exams, eye, vision, glaucoma, cataracts, skin, myopia, regular, macular, degeneration \\
& \textit{[3]} kidney, obesity, hypertension, type, diabetes, sodium, salt, cardiovascular, south, asians \\
\midrule
\textbf{Korean} & 
\textit{[4]} exams, eye, vision, glaucoma, cataracts, skin, myopia, regular, macular, degeneration \\
& \textit{[3]} kidney, obesity, hypertension, type, diabetes, sodium, salt, cardiovascular, south, asians \\
& \textbf{[15]} tested, screened, vaccinated, complications, prevent, hepatitis, necessary, receive, vaccinations, treatment \\
\midrule
\textbf{Taiwanese} & 
\textit{[4]} exams, eye, vision, glaucoma, cataracts, skin, myopia, regular, macular, degeneration \\
& \textit{[3]} kidney, obesity, hypertension, type, diabetes, sodium, salt, cardiovascular, south, asians \\
& \textbf{[25]} stay, date, screenings, reduce, adopt, recommended, uptodate, cancer, healthy, important \\
\midrule
\textbf{Chinese} & 
\textit{[4]} exams, eye, vision, glaucoma, cataracts, skin, myopia, regular, macular, degeneration \\
& \textit{[3]} kidney, obesity, hypertension, type, diabetes, sodium, salt, cardiovascular, south, asians \\
& \textbf{[25]} stay, date, screenings, reduce, adopt, recommended, uptodate, cancer, healthy, important \\
\midrule
\textbf{Bangladeshi} & 
\textbf{[27]} underreporting, anxiety, depression, mental, acculturation, stigmatized, seeking, stigma, discrimination, challenges \\
& \textit{[3]} kidney, obesity, hypertension, type, diabetes, sodium, salt, cardiovascular, south, asians \\
& \textit{[31]} thyroid, disorders, hypothyroidism, digestive, bowel, irritable, syndrome, hyperthyroidism, women, intolerance \\
\midrule
\textbf{Malaysian} & 
\textit{[3]} kidney, obesity, hypertension, type, diabetes, sodium, salt, cardiovascular, south, asians \\
& \textbf{[27]} underreporting, anxiety, depression, mental, acculturation, stigmatized, seeking, stigma, discrimination, challenges \\
& \textbf{[28]} intergenerational, stigmatized, conflicts, acculturation, challenges, mental, unique, discrimination, culture, depression \\
\midrule
\textbf{Indonesian} & 
\textbf{[29]} colorectal, breast, prevalent, lung, cancer, incidence, types, certain, cancers, rates \\
& \textbf{[28]} intergenerational, stigmatized, conflicts, acculturation, challenges, mental, unique, discrimination, culture, depression \\
& \textbf{[27]} underreporting, anxiety, depression, mental, acculturation, stigmatized, seeking, stigma, discrimination, challenges \\
\midrule
\textbf{Indian} &  
\textit{[31]} thyroid, disorders, hypothyroidism, digestive, bowel, irritable, syndrome, hyperthyroidism, women, intolerance \\
& \textbf{[11]} weightbearing, bone, supplements, vitamin, exercises, calcium, consume, adequate, consider, supplementation \\
& \textbf{[2]} vitamin, osteoporosis, deficiency, skin, bones, condition, darker, hormonal, postmenopausal, sunlight \\
\midrule
\textbf{Pakistani} & 
\textbf{[27]} underreporting, anxiety, depression, mental, acculturation, stigmatized, seeking, stigma, discrimination, challenges \\
& \textit{[3]} kidney, obesity, hypertension, type, diabetes, sodium, salt, cardiovascular, south, asians \\
& \textbf{[2]} vitamin, osteoporosis, deficiency, skin, bones, condition, darker, hormonal, postmenopausal, sunlight \\
\midrule
\textbf{Sri Lankan} & 
\textbf{[26]} incidence, cancer, breast, colorectal, stomach, types, prostate, prevalent, certain, detection \\
& \textit{[31]} thyroid, disorders, hypothyroidism, digestive, bowel, irritable, syndrome, hyperthyroidism, women, intolerance \\
& \textbf{[23]} function, kidney, pressure, blood, regularly, changes, monitor, intake, salt, monitoring \\
\midrule
\textbf{Filipino} &  
\textbf{[29]} colorectal, breast, prevalent, lung, cancer, incidence, types, certain, cancers, rates \\
& \textbf{[28]} intergenerational, stigmatized, conflicts, acculturation, challenges, mental, unique, discrimination, culture, depression \\
& \textbf{[25]} stay, date, screenings, reduce, adopt, recommended, uptodate, cancer, healthy, important \\
\bottomrule
\end{tabular}
\vspace{1mm}
\caption{Topic clusters closest to each demographic group, listed in order of relationship strength. Topics that appear uniquely in each group are annotated in bold, while topics that also appear closer to the reference group (`Asian') are annotated in italics. — Group 1}
\label{tab:closest_topic_clusters_R1_G1}
\end{table*}

\begin{table*}[t]
\centering
\begin{tabular}{p{0.15\textwidth}|p{0.8\textwidth}}
\toprule
\textbf{Group} & \textbf{Closest topic clusters} \\
\midrule
\textbf{Vietnamese} \newline
\textbf{Thai} \newline
\textbf{Laotian} \newline
\textbf{Cambodian} \newline
\textbf{Nepalese} \newline
\textbf{Mongolian} \newline
\textbf{Hmong} \newline
\textbf{Burmese} \newline
\textbf{Bhutanese} & 
\textbf{[18]} pain, dental, tobacco, obesity, use, type, cuisine, oral, hypertension, diseases \newline
\textbf{[19]} advocate, resources, healthcare, ensure, seek, needs, providers, important, community, better \newline
\textbf{[21]} advocate, culturally, competent, healthcare, providers, interpretation, background, interpreters, understand, ensure \\
\bottomrule
\end{tabular}
\vspace{1mm}
\caption{Topic clusters closest to each demographic group, listed in order of relationship strength. Topics that appear uniquely in each group are annotated in \textbf{bold}, while topics that also appear closer to the reference group ('Asian') are annotated in \textit{italics}. — Group 2}
\label{tab:closest_topic_clusters_R1_G2}
\end{table*}

To identify the most representative words for each cluster, we first group the topics assigned to each cluster. For each cluster, we calculate the maximum probability for every word across the topics in that cluster, ensuring that the word with the highest probability is retained. This process allows us to rank words by their significance in the cluster. Finally, we select the top 10 words (Tables~\ref{tab:closest_topic_clusters_R1_G1}--\ref{tab:closest_topic_clusters_R1_G3}) based on their maximum probabilities, providing a clear understanding of the most important words characterizing each cluster.

\begin{table*}[t]
\centering
\small
\renewcommand{\arraystretch}{1.2}
\begin{tabular}{p{0.15\textwidth}|p{0.8\textwidth}}
\toprule
\textbf{Group} & \textbf{Closest topic clusters} \\
\midrule
\textbf{Japanese}, \textbf{Okinawan} &
\textbf{[30]} concerned, include, obesity, american, hypertension, alzheimers, pressure, issues, important, blood \newline
\textbf{[11]} weightbearing, bone, supplements, vitamin, exercises, calcium, consume, adequate, consider, supplementation \newline
\textit{[4]} exams, eye, vision, glaucoma, cataracts, skin, myopia, regular, macular, degeneration \\
\bottomrule
\end{tabular}
\vspace{1mm}
\caption{Topic clusters closest to each demographic group, listed in order of relationship strength. Topics that appear uniquely in each group are annotated in \textbf{bold}, while topics that also appear closer to the reference group ('Asian') are annotated in \textit{italics}. — Group 3}
\label{tab:closest_topic_clusters_R1_G3}
\end{table*}

\paragraph{Jensen--Shannon divergence (JSD)} In our analysis, we calculate the JSD for each pair of demographic groups using their 
respective cluster proportion vectors. Missing clusters in any group are set to zero to ensure 
consistent vector lengths. The JSD between two probability distributions \( p \) and \( q \) is defined as:
\[
\text{JS}(p, q) = \frac{1}{2} \left( D_{\text{KL}}(p \, || \, m) + D_{\text{KL}}(q \, || \, m) \right)
\]
where \( m \) is the midpoint distribution:
\[
m = \frac{1}{2}(p + q)
\]
and the Kullback--Leibler (KL) divergence is defined as:
\[
D_{\text{KL}}(p \, || \, q) = \sum_{x} p(x) \log \frac{p(x)}{q(x)}.
\]
In our context, the probability $p$ and $q$ represent the proportion of topics in each cluster for the two demographic groups (e.g., the $i$-th element of $p$ denotes the proportion of the $i$-th topic). The divergence captures the dissimilarity between their 
topic distributions, providing a quantitative measure for comparing the groups.

\end{document}